\documentclass[11pt]{article}
\pdfoutput=1
\usepackage{amsmath,amssymb,color,cite}
\usepackage{slashed}
\usepackage{graphicx}
\usepackage{amsfonts}
\usepackage{enumerate}
\usepackage{hyperref}
\usepackage{bbm}
\usepackage{nicefrac}
\usepackage[all]{xy}
\usepackage{graphicx}
\usepackage{booktabs}
\usepackage{epstopdf}
\usepackage{simplewick}
\usepackage{amsfonts}
\newcommand{\hoch}[1]{$\, ^{#1}$}

\makeatletter
\@addtoreset{equation}{section}
\makeatother

\newcommand{\be}{\begin{equation}}
\newcommand{\ee}{\end{equation}}
\newcommand{\bea}{\setlength\arraycolsep{2pt} \begin{eqnarray}}
\newcommand{\eea}{\end{eqnarray}}

\def\im{{{\rm i}}}

\begin{document}

\vspace{25pt}
\begin{center}
{\Large On the $\phi^3$ Theory Above Six Dimensions{\bf }
}

\vspace{30pt}

{\Large
 Junchen Rong\hoch{1},  Jierong Zhu\hoch{2},
}

\vspace{10pt}

\hoch{1} {\it DESY Hamburg, Theory Group, Notkestraße 85, D-22607 Hamburg, Germany
}\\
\hoch{2} {\it CAS Key Laboratory of Theoretical Physics, Institute of Theoretical Physics,\\ Chinese Academy of Sciences, Beijing 100190, China}
\vspace{10pt}

\vspace{20pt}

ABSTRACT
\end{center}
\vspace{15pt}
{We study the scalar $\phi^3$ theory above six dimensions. The beta function $\beta(g)=-\epsilon g-\frac{3}{4}g^3$ in $d=6-2\epsilon$ dimensions has a UV fixed point when $\epsilon<0$.  Like the $O(N)$ vector models above four dimensions, such a fixed point observed perturbatively in fact corresponds to a pair of complex CFTs separated by a branch cut. Using both the numerical bootstrap method and Gliozzi's fusion rule truncation method, we argue that the fixed points of the $\phi^3$ theory above six dimensions exist.}

\thispagestyle{empty}

\pagebreak
\voffset=-40pt
\setcounter{page}{1}

%\tableofcontents

%\addtocontents{toc}{\protect\setcounter{tocdepth}{2}}

%%%%%%%%%%%%%%%%%%%%%%%%%%%%%%%%%%%%%%%%
\section{Introduction}
Consider a scalar field theory with the following Lagrangian 
\be\label{singlescala}
\mathcal{L}= \frac{1}{2}\partial_{\mu}\phi  \partial^{\mu}\phi +\frac{1}{3!}g \phi^3.
\ee
The one loop beta function in $6-2\epsilon$ dimensions has the following form 
\be
\beta(g)=-\epsilon g-\frac{3}{4}g^3+\ldots.
\ee
Below six dimensions, the theory has a fixed point at the purely imaginary coupling $g_{*}=\im \frac{2\sqrt{3}}{3} \sqrt{\epsilon}$. This fixed point was used by Michael  Fisher to study the Lee-Yang edge singularity \cite{Fisher:1978pf}. Formally, if one were to take $\epsilon$ to be negative (i.e., $d>6$), one finds a fixed point. It can be shown that $\Delta_{\phi}$ is greater that the unitarity bound. The fixed point is perturbatively unitary. The existence of such a non-renormalizable fixed point, at least perturbatively, was noticed in the 70's\cite{Mack:1973kaa}, and the leading correction to the scaling dimension $\Delta_{\phi}$ was calculated by solving the Migdal-Polyakov bootstrap equations \cite{Migdal:1971xh,Polyakov:1970xd}.

The idea behind the modern bootstrap method also dates back to the 70's \cite{Polyakov:1974gs,Ferrara:1973yt}. It was later applied to the two dimensional conformal field theories in the famous work of Belavin, Polyakov and Zamolodchikov \cite{Belavin:1984vu}, where the two dimensional minimal models were solved. It was not until 2008 that some important progress was made in applying the conformal bootstrap method to CFTs in $d>2$, with the help of a computer \cite{Rattazzi:2008pe}.  After that, the numerical bootstrap became an important method to study conformal field theories in various dimensions, see \cite{Poland:2018epd} for a recent review. In some key examples, the critical exponents of certain models calculated using the numerical bootstrap method were more precise than the results from Monte-Carlo simulations \cite{ElShowk:2012ht,El-Showk:2014dwa,Kos:2015mba,Kos:2014bka,Kos:2016ysd,Chester:2019ifh}.

Since the $\phi^3$ interaction in \eqref{singlescala} is irrelevant in $d>6$, the interacting fixed point is in the Ultra-Violet. The coupling constants of the irrelevant terms are growing as the renormalization group flow approaches the interacting fixed point. Higher weight terms such as $\phi^4, \phi^5 \ldots$ should also play a rule in the renormalization.  We do not really know whether we can trust the naive $\epsilon$ expansion based on a single $\phi^3$ interaction term at large and negative $\epsilon$. It will therefore be interesting to study the fixed point using other methods. Another important remark is that the scalar theory with cubic interaction is intrinsically non-unitary, due to the metastability of the $\phi^3$ potential. We expect the scaling dimension $\Delta_{\phi}$ and some OPE coefficients to have small imaginary parts as $d\rightarrow 6_+$. Similar perturbatively unitary fixed points were shown to exist for $O(N)$ vector models in $d=4+\epsilon$ \cite{McKane:1984eq}, and in $4<d<6$ for large enough $N$  \cite{Fei:2014yja,Fei:2014xta,Gracey:2015tta}. In both cases, using the instanton method, it was shown that both the scaling dimension $\Delta_{\phi}$ and some OPE coefficients receive small imaginary corrections \cite{McKane:1984eq,Giombi:2019upv}. At large enough $N$, the non-perturbative imaginary parts are numerically small, so that these fixed points appear in the corresponding numerical bootstrap results \cite{Bae:2014hia,Chester:2014gqa,Li:2016wdp}.  Borrowing the intuition from these previous works, we expect that the $\Delta_{\phi}$ and  OPEs of the $\phi^3$ fixed point above six dimensions to also develop small imaginary parts as $d\rightarrow 6_+$.

In this work, we use two different methods to study the fixed point of the $\phi^3$ theory. Since the conformal blocks can be defined in non-integer dimensions \cite{El-Showk:2013nia}, one can use the numerical bootstrap method to study the $\phi^3$ theory in $6<d<7$ \footnote{Notice that according to the work of \cite{Hogervorst:2014rta,Hogervorst:2015akt}, the $\phi^3$ theory in non-integer dimensions are inherently non-unitary, due to the existence of the so-called ``evanescent operators'' which decouple in integer dimensions. These operators have large scaling dimensions. We can safely neglect them when doing numerical bootstrap, the same strategy was adopted in \cite{El-Showk:2013nia}.  }. We compare the perturbative $\epsilon$-expansion results in \cite{Gracey:2015tta} with the results from the numerical bootstrap. In close to six dimensions, the boundary of allowed region in the $\left(\Delta_{\phi},\Delta'\right)$ plane shows a sharp cliff precisely at the value of $\Delta_{\phi}$ predicted by the $\epsilon$-expansion, see Fig. \ref{numericalB}. Here $\Delta'$ denotes the scaling dimension of the second primary operator appearing in the $\phi\times \phi$ OPE. Notice that the first primary that appears in the $\phi\times \phi$ OPE is $\phi$ itself. As the space-time dimension $d$ increases, the cliff suddenly disappears. This leads us to the conjecture that one of the $\lambda_{\phi\phi\mathcal{O}}^2$ (square of operator product expansion coefficient), after neglecting the instantons' effect, changes sign at this dimension. After the numerical bootstrap study, we then use Gliozzi's fusion rule truncation bootstrap method \cite{Gliozzi:2013ysa,Gliozzi:2014jsa} to study the $\phi^3$ fixed point by focusing on the fusion rule $\left[ \Delta_\phi\right] \times\left[ \Delta_\phi\right] =1+\left[ \Delta_\phi\right] +\left[ \Delta_T\right] +\left[ \Delta_4\right] $. The calculation is precisely the same as in \cite{Gliozzi:2013ysa}, except that we now set $d>6$.  We show that $\lambda^2_{\phi\phi T}$ is negative above a certain $d_c$ which is non-integer.

This paper is organized as follows. In Section \ref{bootstrap}, we review the standard numerical bootstrap method and give the results for the $\phi^3$ theory in $6<d<7$, the maximum number of derivatives used for the numerical bootstrap is set to be either $\Lambda=23$ or $35$. We then compare the results with the anomalous dimension of $\phi$  calculated perturbatively at four loops. In Section \ref{truncation}, we use Gliozzi's fusion rule truncation bootstrap to study the same theory, and compare the results with that of the numerical bootstrap. Some discussions on the implication of our results are given in Section \ref{disscuss}. 

\section{Numerical bootstrap}\label{bootstrap}
Due to the conformal symmetry, the four point function of four identical scalars is fixed to be
\be
\langle\phi(x_1)\phi(x_2)\phi(x_3)\phi(x_4)\rangle= \frac{g(u,v)}{|x_{12}|^{2\Delta_{\phi}}|x_{34}|^{2\Delta_{\phi}}},
\ee
where the unfixed function $g(u,v)$ depends on the cross ratios $u=\frac{x_{12}^2 x_{34}^2}{x_{13}^2 x_{24}^2}$ and $v=\frac{x_{14}^2 x_{23}^2}{x_{13}^2 x_{24}^2}$. The function $g(u,v)$ admits the following conformal block expansion
\be\label{singleblockexpansion}
g(u,v)=1+\sum_{\mathcal{O}}\lambda_{\mathcal{O}}^2 G_{\Delta,l}(u,v).
\ee
For a physical CFT, such a series expansion is convergent in a certain region of the $u$-$v$ plane, and the four point function is crossing symmetric:
\be\label{crossingsym}
\langle
\contraction{}{\phi_i(x_1)}{}{\phi_j(x_2)}
\phi_i(x_1)\phi_j(x_2)
\contraction{}{\phi_k(x_3)}{}{\phi_{l}(x_4)}
\phi_k(x_3)\phi_{l}(x_4)
\rangle=\langle
\contraction{}{\phi_i(x_1)}{\phi_j(x_2)\phi_k(x_3)}{\phi_{l}(x_4)}
\contraction[2ex]{\phi_i(x_1)}{\phi_i(x_2)}{}{\phi_{l}(x_3)}
\phi_i(x_1)\phi_j(x_2)\phi_k(x_3)\phi_{l}(x_4)
\rangle.
\ee
The lines shown in the above equation indicate how the operator product expansion is performed. Equation \eqref{crossingsym} gives us the following crossing equation \cite{Rattazzi:2008pe},
\bea\label{crossingequation}
&&\sum_{\mathcal O}\lambda_{\phi\phi\mathcal O}^2 F_{\Delta,l}=0,\nonumber\\
\text{with}\quad&& F_{\Delta,l}=v^{\Delta_{\phi}}G_{\Delta,l}(u,v)-u^{\Delta_{\phi}}G_{\Delta,l}(v,u).
\eea
In a bootstrap setup, we numerically search for a linear functional $\alpha$ such that 
\bea\label{crscondition}
&&\alpha(F_{0,0})=1, \nonumber\\
&&\alpha(F_{\Delta,l})\geq0 \quad \text{for } \Delta=\Delta_{\phi},\quad  l=0, \nonumber\\
&&\alpha(F_{\Delta,l})\geq0 \quad \text{for } \Delta\geq \Delta', \quad l=0, \nonumber\\
&&\alpha(F_{\Delta,l})\geq0 \quad \text{for } \Delta\geq d-2+l, \quad l=2,4,6\ldots.
\eea
The above conditions are very similar to the conditions that were used to bootstrap the Ising model. The critical difference is that we allow the operator $\phi$ to appear in the OPE $\phi\times\phi$. Notice that for the $\phi^3$ theory, the three point function $\langle\phi(x_1)\phi(x_2)\phi(x_3)\rangle$ is non-vanishing. For a unitary CFT, the OPE coefficients are real numbers, and hence $\lambda_{\phi\phi\mathcal{O}}^2$ is positive. For a chosen pair of $\{\Delta_{\phi},\Delta'\}$, if such a linear function $\alpha$ is found, then there is no way the crossing equation \eqref{crossingequation} can be satisfied. This can be seen by applying $\alpha$ to both sides of the equation. Physically this means that there exist no unitary CFTs with $\phi$ being the only scalar primary operator whose dimension is lower that $\Delta'$.  

\begin{figure}[h!]
\begin{center}
\includegraphics[width=0.49\textwidth]{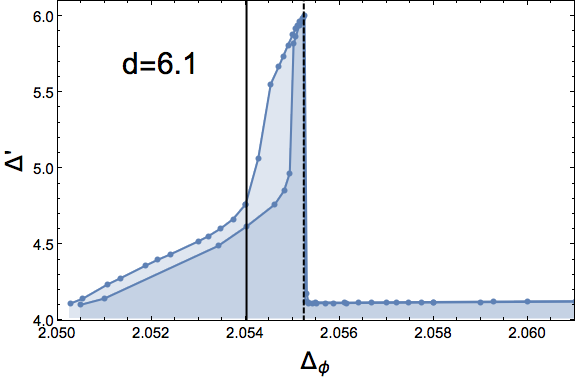}
\includegraphics[width=0.49\textwidth]{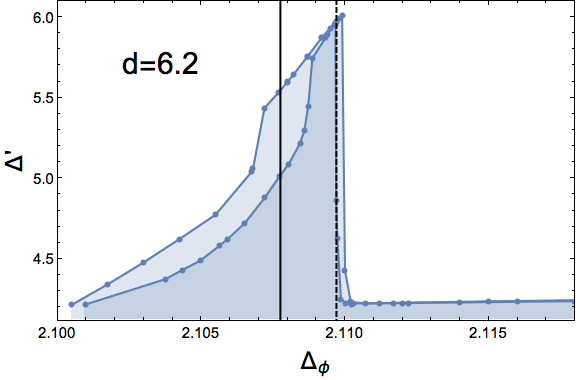}\\
\includegraphics[width=0.49\textwidth]{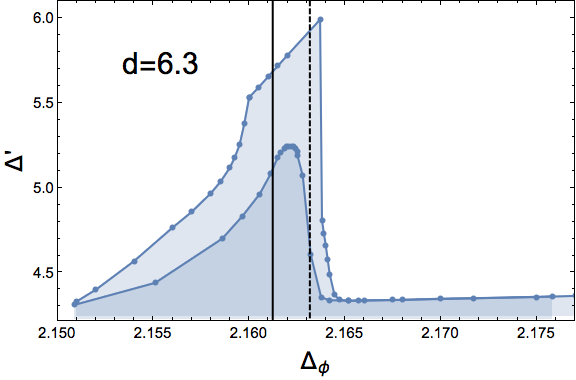}
\includegraphics[width=0.49\textwidth]{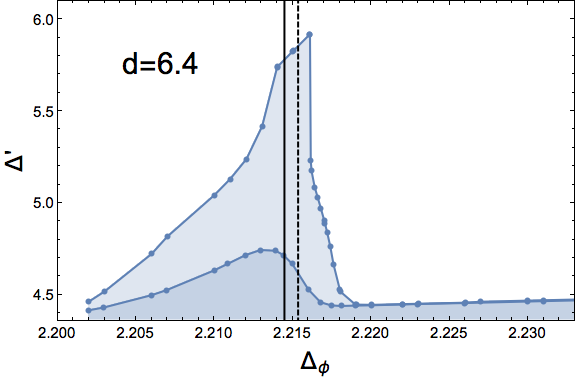}\\
\includegraphics[width=0.49\textwidth]{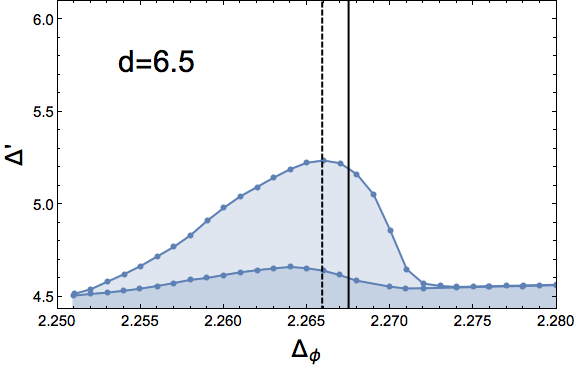}
\includegraphics[width=0.49\textwidth]{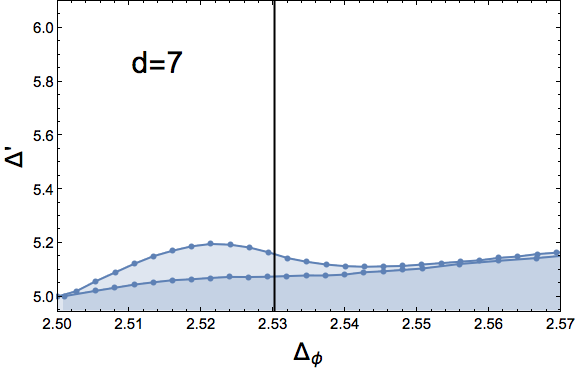}
\end{center}
\caption{The numerical bootstrap results in $6<d<7$. The weaker/stronger bound corresponds to $\Lambda=23/35$ respectively. The dashed lines are the predictions of $\Delta_{\phi}$ from the $\epsilon$-expansion. The solid lines correspond to $\Delta_{\phi}$ calculated using Gliozzi's fusion rule truncation method. \label{numericalB}}
\end{figure}

The problem of searching for such a linear functional can be translated into a semi-definite programming problem. In this work, we use the ``SDPB" solver, which was designed to study the conformal bootstrap problems \cite{Simmons-Duffin:2015qma}. The numerical studies in this paper are performed using two sets of parameters. In the first case, the maximum number of derivatives is chosen to be $\Lambda=23$  and the range of spins is set to be $l \in\{0,\ldots 26\}\cup\{49,50\}$. In the second case, the maximum number of derivatives is $\Lambda=35$ and the range of spins is $l \in\{0,\ldots 44\}\cup\{47,48,51,52,55,56,59,60,63,64,67,68\}$ \footnote{Such a choice of spins is inherited from the seminal work of \cite{Simmons-Duffin:2015qma}. The result will not change if we include all the spins $l \in {0,\ldots 50/68}$ for $\Lambda=23/35$. Neglecting certain higher spin constrains helps us save a bit of computational time.}. The results of the numerical bootstrap are summarised in Fig. \ref{numericalB}.

The crossing equation has a set of ubiquitous solutions in any space-time dimensions, which are the so called generalised free field theories. A generalised free field theory is equivalent to a free scalar propagating in $AdS_{d+1}$, see for example \cite{Heemskerk:2009pn}. The operators appearing in the $\phi\times \phi$ OPE have scaling dimensions given by
\be
\Delta=2\Delta_{\phi}+l,\quad \text{ with } l\in \text{even}.
\ee
The scaling dimension $\Delta_{\phi}$ can be a generic real number. Strictly speaking, it is not a full-fledged conformal field theory since there is no conserved spin-2 current, therefore no stress-energy tensor. From the numerical bootstrap point of view, such CFTs satisfy both the crossing symmetry and the unitarity constraints, and therefore should fall into the allowed region of the bootstrap plot. One may worry about the fact that the operator $\phi$ does not appear in the $\phi\times \phi$ OPE in generalized free theories. Notice that even though the conditions \eqref{crscondition} allow $\phi$ to appear, its presence is however not required. Generalized free theories are compatible with such conditions. They correspond to the baseline $\Delta'=\Delta_{\phi^2}=2\Delta_{\phi}$ in the numerical bootstrap curve. Above this baseline, we observe some interesting excess. When $d$ is close to 6, the right hand side of the excess area is a sharp cliff. The excess area becomes a much smoother mountain as $d$ increase. The mountain becomes even more flat as $d$ approaches 7. 
Close to $d=6$, we expect that the perturbation calculation to give a reasonably accurate prediction of the scaling dimension $\Delta_\phi$. We quote here the formula in \cite{Gracey:2015tta} at four loops:
\bea\label{singlescalarloop}
\Delta_{\phi}&=&2-\frac{10 \epsilon }{9}-\frac{86 \epsilon ^2}{729}+\frac{(15552 \zeta (3)-8375) \epsilon ^3}{59049}\nonumber\\
&&+\frac{(-2783808 \zeta (3)+3779136 \zeta (4)-2799360 \zeta (5)-3883409) \epsilon ^4}{2\ 4782969}+\mathcal{O}(\epsilon^4).
\eea
The dashed lines in Fig. \ref{numericalB} correspond to the predictions of the $\epsilon$-expansion, where we have used the Pad\'e$_{[1,3]}$ method to resum the series. At close to 6d,  the sharp cliff is precisely located at the value predicted by the $\epsilon$-expansion method. Such non-smoothness usually indicates a conformal field theory at the non-smooth point. A famous example is the kink observed in the three dimensional numerical bootstrap bound, which corresponds the Ising model \cite{ElShowk:2012ht}.  The existence of such a cliff gives us confidence that indeed the fixed point of the $\phi^3$ theory exists above six dimensions. At close to six dimensions, the appearance of the cliff could also be understood from the equation of motion of the $\phi^3$ theory, $\Box \phi=\frac{1}{2}g\phi^2$. Clearly the operator $\phi^2$ is now a conformal descendant of $\phi$ and the next to leading conformal primary scalar operator is  $\phi^3$, whose scaling dimension should be much bigger than $2\Delta_{\phi}$. The critical $d_c$ at which the cliff disappears depends on $\Lambda$---the number of derivatives used in the numerical bootstrap. At $\Lambda=23$, we get $6.4< d_c<6.5$. At $\Lambda=35$, we get $6.2< d_c<6.3$. The disappearance of the bootstrap cliff is affected by two factors. First, as was mentioned in the introduction, the $\phi^3$ theory above six dimension is non-unitary due to instantons. As $d$ increase, the imaginary parts of $\Delta_{\phi}$ and some OPE coefficients may become bigger, so that they are not negligible anymore. Second, even if it is safe to neglect the instantons' effect, the OPE$^2$ of some low lying operators may change sign at a certain critical space-time dimension. We will show in next section that $\lambda_{\phi\phi T}^2$ does change sign at $d\approx 6.43$, with $T$ being the stress-energy tensor.

\section{Gliozzi's Fusion Rule Truncation Bootstrap}\label{truncation}
In this section, we use the method developed in \cite{Gliozzi:2013ysa} to study the $\phi^3$ theory above $d=6$. The fusion rule truncation method allows us to extract information on the operators with low scaling dimensions in the $\phi\times\phi$ OPE.  The calculation is precisely the same as in \cite{Gliozzi:2013ysa}, except that we now set $d>6$. One advantage of Gliozzi's method is that it works also for non-unitary CFTs, as long as the theory can be approximated by a truncated fusion rule. We follow the steps in\cite{Gliozzi:2013ysa} and define 
\be
f^{(m,n)}_{\Delta,l}=\partial^m_a\partial^n_b\frac{v^{\Delta_{\phi}} G_\beta(u,v)-u^{\Delta_{\phi}} G_\beta(v,u)}{u^{\Delta_{\phi}}-v^{\Delta_{\phi}}}|_{a,b=1,0},
\ee
where two variables $a,b$ are related to the more familiar cross ratios $z$ and $\overline{z}$ by $z=(a+\sqrt{b})/2, \overline{z}=(a-\sqrt{b})/2$.

The crossing equation $\eqref{crossingequation}$ leads to the following infinite number of homogeneous equations
\be\label{homogeneouseq}
\sum_{\Delta,l}\lambda^2_{\Delta,l}f^{(2m,n)}_{\Delta,l}=0\qquad(m\geq0, n\geq0, m+n>0),
\ee
and a normalization condition 
\be\label{inhomogeneouseq}
\sum_{\Delta,l}\lambda^2_{\Delta,l}f^{(0,0)}_{\Delta,l}=1.
\ee

One may attempt to solve \eqref{homogeneouseq} in the following way. Truncate  \eqref{homogeneouseq} to the first $N$ operators and the first $M$ equations, with $M>N$. The left hand side of \eqref{homogeneouseq} therefore becomes a $M\times N$ dimensional matrix times a $N$-dimensional vector. If the equations have a non-zero solution, this means that any $N\times N$ minors of the matrix should have vanishing determinants. Such CFTs are called ``truncable'' in \cite{Gliozzi:2014jsa}.

\begin{figure}[h!]
\begin{center}
\includegraphics[width=0.49\textwidth]{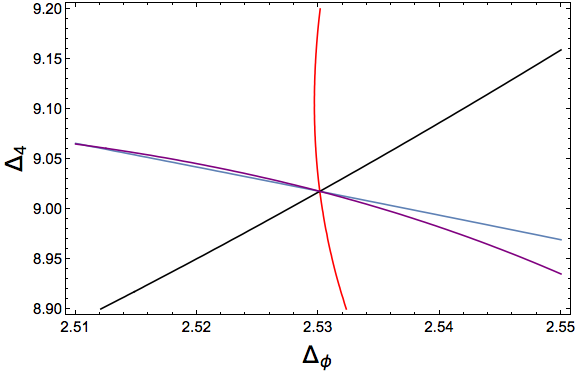}
\end{center}
\caption{Determinants of the $3\times3$ minors of the truncated crossing equations at $d=7$.}\label{det}
\end{figure}

As in \cite{Gliozzi:2013ysa}, we truncate the fusion rule of $\phi^3$ theory to be 
\be\label{Tfusion}
[\Delta_{\phi}]\times [\Delta_{\phi}]=1+[\Delta_{\phi}]+[D,2]+[\Delta_{4},4]+\ldots.
\ee 
Here $[D,2]$ denotes the stress energy tensor, and $[\Delta_4,4]$ denotes a spin-4 operator with scaling dimension $\Delta_4$. We will also restrict our attention to the four equations in \eqref{homogeneouseq}, with $(m,n)=(1,0),(2,0),(0,1),(0,2)$, as in \cite{Gliozzi:2013ysa}.  In this case we have $\frac{4!}{3!}=4$ possible $3\times 3$ minors. The determinants of the $3\times3$ minors at $d=7$ are shown in Fig. \ref{det}.
One can see that they intersect at an unique point, which corresponds to the $\phi^3$ theory. Setting $\Delta_{\phi}$ and $\Delta_4$ to be at the point where the determinants of the minors intersect, we can now solve \eqref{homogeneouseq} with a non-zero vector.  We then use \eqref{inhomogeneouseq} to fix its normalization. It is important to note that the solutions we found have purely real scaling dimensions and the squares of the OPE coefficients are also real. This means that we are neglecting the imaginary parts due to instantons.

\begin{figure}[h!]
\begin{center}
\includegraphics[width=0.49\textwidth]{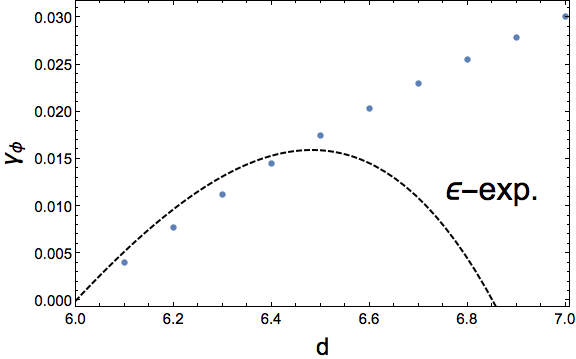}
\includegraphics[width=0.49\textwidth]{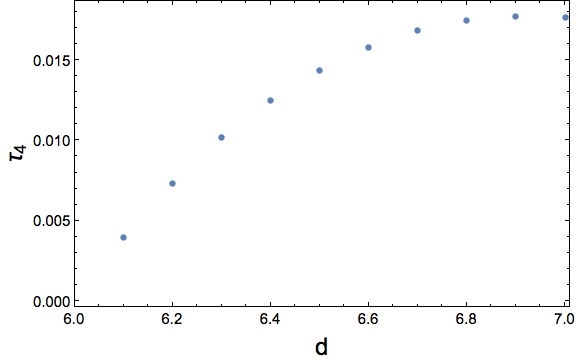}\\ 
\vspace{1cm}
\includegraphics[width=0.49\textwidth]{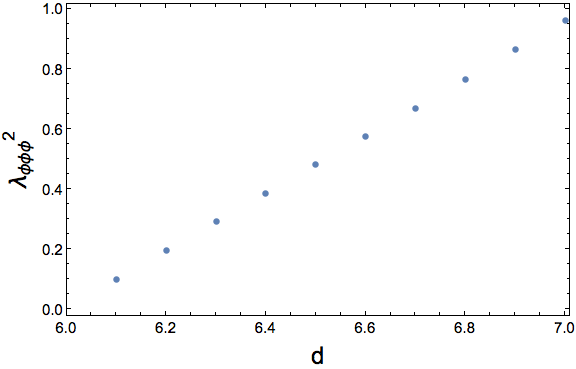}
\includegraphics[width=0.49\textwidth]{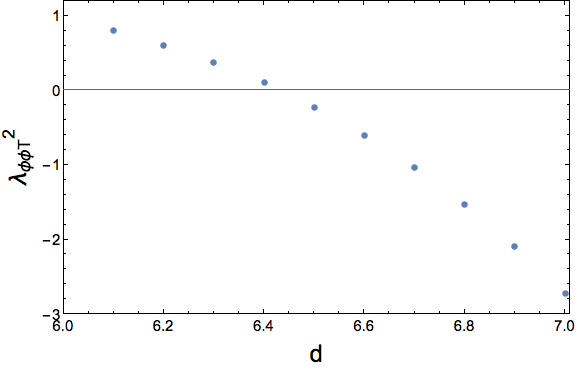}
\end{center}
\caption{Scaling dimensions and OPE coefficients of the $\phi^3$ fixed point calculated using Gliozzi's fusion rule truncation method. We use the definition that $\gamma_{\phi}=\Delta_{\phi}-\frac{d-2}{2}$ and $\tau_4=\Delta_4-(d+2)$. Notice that $\lambda_{\phi\phi T}^2$ changes sign at $d_c\approx 6.43$. In our convention, the central charge is related to OPE coefficient as $c/c_{\text{free boson}}=\frac{\Delta_{\phi}^2}{\lambda_{\phi\phi T}^2}$. The dashed line in the $\Delta_{\phi}$ plot is the Pad\'e$_{[1,3]}$ re-summation of the four loop $\epsilon$-expansion series.\label{GFTB}}
\end{figure}

It is not clear why such a truncated fusion rule should allow us to solve the $\phi^3$ theory. It was shown that the same truncated fusion rule can be used to calculate the critical exponents of the Lee-Yang edge singularities below 6 dimensions\cite{Gliozzi:2013ysa}. A cross check with  the results of the numerical bootstrap and the $\epsilon$-expansion is helpful. The solutions in $6<d<7$ are summarised in Fig. \ref{GFTB}. We have also marked the values of $\Delta_{\phi}$ in Fig. \ref{numericalB} using solid lines.
For all these solutions, we have checked that the three primary operators (and the identity operator) kept in the fusion rule \eqref{Tfusion} are enough to solve the $M=4$ homogenous equation in \eqref{homogeneouseq} with $(m,n)=(1,0),(2,0),(0,1),(0,2)$. Notice that $\lambda_{\phi\phi T}^2$ changes sign at $d_c\approx 6.43$, which is possibly related to the disappearance of the cliff in the bootstrap results. Also, comparing to the $\epsilon$-expansion, the fusion rule truncation method is less accurate as $d\rightarrow6$. This is because the fusion rule of the $\phi^3$ theory becomes more like a free theory fusion rule, and the truncation to three primaries is not valid. Notice that when doing the numerical bootstrap with a smaller number of derivatives ($\Lambda=23$), we do observe a bump of the bootstrap curve at $d=7$, which is roughly located at the $\Delta_{\phi}$ predicted by the fusion rule truncation method. This seems to suggest that the non-unitary solutions to the crossing equation can affect the bootstrap curve in low derivatives.  

The solutions to the truncated bootstrap equations can be continued to even higher space-time dimensions. We have checked that they exist at $d=8,9$ and $10$, which are summarised in Table \ref{higherD}. We however emphasize here that there is no reason to assume the instantons' effects to be small. We are not sure whether the solutions to the truncated fusion rule equations are good approximations of the fixed points of the $\phi^3$ theories in the corresponding space-time dimensions. In fact, it would also be interesting to search for solutions with complex operator scaling dimensions \footnote{We thank the referee for suggesting this.}. To interpret them as the $\phi^3$ theory, their imaginary parts shall become negligible when continued to $d=6$. We performed a preliminary search but did not find such solutions. We leave a more detailed study as a future problem.  
\begin{table}[h]
\centering
\begin{tabular}{|c|c|c|c|c|c|}
 \hline \hline
$d$ &  8 & 9& 10 \\\hline
$\gamma_{\phi}$&0.0478&0.0611&0.0729\\\hline
$\tau_4$ &0.0008&-0.0421&-0.1114\\\hline
\end{tabular}
\caption{Solutions to the truncated crossing equations at $d>7$.}\label{higherD}
\end{table}

\section{Discussion}\label{disscuss}
\begin{figure}[h!]
\begin{center}
\includegraphics[width=1\textwidth]{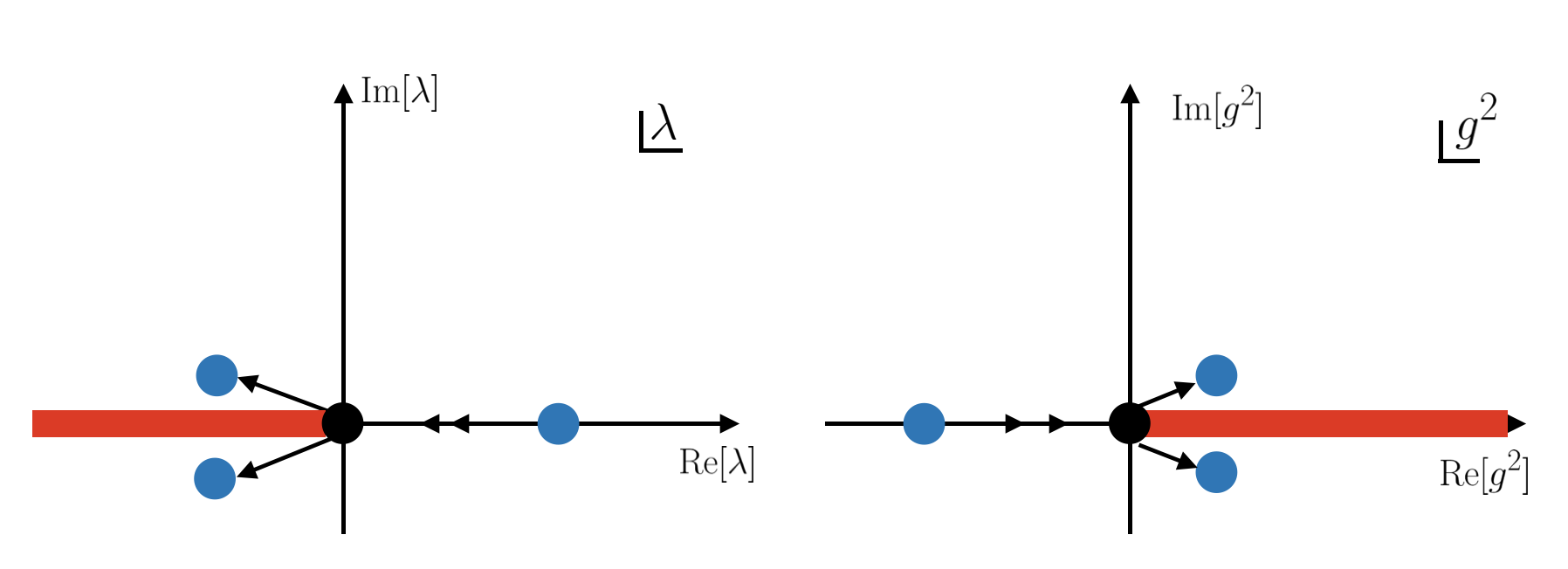}
\end{center}
\caption{Evolution of the fixed points of the $\lambda \phi^4$ theory and the $g \phi^3$ theory as the space time dimension $d$ increases. The Wilson-Fisher/Lee-Yang fixed point (the blue dots) hits the free theory (the black dot) at $d=4/6$, after which it bifurcates into a pair of complex CFTs. Note that the free theory lives at the tip of a branch cut (the red line). In the case of the $\lambda \phi^4$ theory, the branch cut lies along the negative $\lambda$ axis, while in the case of the $g \phi^3$ theory, the branch cut lies along the positive $g^2$ axis. }\label{CFTbifurcate}
\end{figure}
As mentioned in the introduction, we expect the critical exponents and OPE coefficients of the fixed point of the $\phi^3$ theory above six dimensions to have small imaginary parts due to the instantons. This means that we have a pair of CFTs with conjugate spectrum and OPE coefficients. Such CFTs with operators of complex scaling dimensions were named ``complex CFTs'' and were shown to be related to the  ``walking''  of RG flows recently in \cite{Gorbenko:2018ncu}. It would be interesting to look into the details of how the pair of CFTs are created. Let's start with the $\lambda \phi^4$ theory, a scalar theory with $\mathcal{L}_{int}=\frac{\lambda}{4!}\phi^4$ interaction. The leading terms of the beta function in $d=4-\epsilon$ is 
\be
\beta (\lambda)=-\lambda g+3 \lambda^2+\ldots
\ee
The coupling at the Wilson-Fisher fixed point is $\lambda_{*}=\frac{\epsilon}{3}$. As we vary the space time dimension from $d<4$ to $d>4$, the Wilson-Fisher fixed point (as indicated by the blue dots in Figure \ref{CFTbifurcate}) hits the free theory fixed point (as indicated by the black dots in Figure \ref{CFTbifurcate}) at $d=4$. Notice that due to the instability of the potential at negative coupling, the free theory in fact lives at the tip of a branch cut. Such a branch cut was argued to exist in QED by Dyson \cite{Dyson:1952tj}, and was made rigorous in \cite{Simon:1970mc} for the one dimensional $\phi^4$ theory, which reduces to a quantum mechanical system. It may be safe to expect a similar cut in higher dimensional theories. In \cite{McKane:1984eq}, the critical exponents of the $\lambda \phi^4$ theory at $d=4+\epsilon$ was shown to have small imaginary parts, and the cut plays an essential role in the calculation. After the Wilson-Fisher fixed point collides with the free theory fixed point at $d=4$, it bifurcates into a pair of complex CFTs with conjugate coupling $\lambda$. 

The creation of the complex CFTs in $g\phi^3$ theory could be understood in a similar manner. In the corresponding one dimensional quantum mechanical system, it was shown in \cite{Vainshtein:1964dw} that such a theory has a branch cut along the real and positive $g^2$ axis. We  can borrow this one dimensional intuition to think about the higher dimensional $g\phi^3$ theory. As $d$ increases, the Lee-Yang edge singularity, which lives on the negative $g^2$ axis, hits the free theory at $d=6$. Due to the branch cut, it then bifurcates into a pair of complex CFTs.  When $d\gg6$, they might be far away from the real axis, it is however hard to imagine that they suddenly disappear. So we expect such a pair of complex CFTs to exist even in higher dimensions.

Similar complex CFTs should also exist in gauge theories. The lower end of the conformal window of the supersymmetric QCD, $3N_c>N_f>\frac{3}{2}N_c$, can also be understood as an interacting fixed point hitting the free theory \cite{Kaplan:2009kr,Gorbenko:2018ncu}. The upper end of the conformal window is fixed by requiring the theory to be asymptotically free. According to the Seiberg duality, we know that the theory has a dual description in terms of a magnetic gauge theory. Even though the electric theory becomes strongly coupled as $N_f\rightarrow \frac{3}{2}N_c$, the dual magnetic gauge theory is weakly coupled. We expect a branch cut to exist when the magnetic gauge theory has negative $g_{m}^2$. As we lower $N_f$, the (super) Banks-Zaks fixed point hits the magnetic free gauge theory and then bifurcates into a pair of complex CFTs.  There exist another ubiquitous phenomenon through which a pair of CFTs could go into the complex plane \cite{Dymarsky:2005uh,Kaplan:2009kr,Fei:2014yja,Pomoni:2008de,Giombi:2015haa,Gorbenko:2018ncu}. In that case, two interacting CFTs collide and then becomes a pair of conjugate complex CFTs. This scenario is different from the scenario when an interacting CFT hits a branch point CFT and bifurcates.  One difference, for example, is that in the later case, after the collision, the branch point CFT is still present. Thus there are three fixed points in the complex plane. As was already mentioned, the lower end of the conformal window of the super QCD terminates because the (super) Banks-Zaks fixed point hits the free theory.  The lower end of the conformal window of the non-supersymmetric QCD, however, may be described by interacting CFTs' merging. This difference was made clear in \cite{Kaplan:2009kr,Gorbenko:2018ncu}. 

Notice that in all the examples mentioned above, the fixed points at the tip of the branch cuts are essentially free theories. It would be interesting to understand whether a branch point CFT can be genuinely interacting. Also, there are other models in even higher dimensions~\cite{Gracey:2015xmw,Gliozzi:2016ysv,Gliozzi:2017hni} which were argued to flow to fixed points in $d>6$, it would be interesting to see whether they fit in the frame of complex CFTs.

\vskip 0.4 in
{\noindent\large  \bf Acknowledgments}
\vskip 0.in
%%%%%%%%%%%%%%%%%%%%%%%%%%%%%%%%%%%%%%%
We are grateful to Ning Su and Nakwoo Kim for helpful discussion and comments. J. R. would like to thank the hospitality of the Asia Pacific Center for Theoretical Physics where the early stage of the work was finished. The numerics is solved using SDPB program\cite{Simmons-Duffin:2015qma} and simpleboot (https://gitlab.com/bootstrapcollaboration/simpleboot). Z.R. was supported by in part by the Thousand Young Talents program of Song He and the Peng Huanwu center under Grant No. 11747601. The work of J.R. is supported by the DFG through the Emmy Noether research group “The Conformal Bootstrap
Program” project number 400570283.


\begin{thebibliography}{99}
%\cite{Fisher:1978pf}
\bibitem{Fisher:1978pf} 
  M.~E.~Fisher,
  ``Yang-Lee Edge Singularity and phi**3 Field Theory,''
  Phys.\ Rev.\ Lett.\  {\bf 40}, 1610 (1978).
  doi:10.1103/PhysRevLett.40.1610
  %%CITATION = doi:10.1103/PhysRevLett.40.1610;%%
  %136 citations counted in INSPIRE as of 20 Aug 2017
%\cite{Gliozzi:2013ysa}
%\cite{ElShowk:2012ht}

%\cite{Mack:1973kaa}
\bibitem{Mack:1973kaa} 
  G.~Mack,
  ``Conformal invariance and short distance behavior in quantum field theory,''
  Lect.\ Notes Phys.\  {\bf 17}, 300 (1973).
  doi:10.1007/BFb0017087
  %%CITATION = doi:10.1007/BFb0017087;%%
  %5 citations counted in INSPIRE as of 21 Jan 2020

%\cite{Migdal:1971xh}
\bibitem{Migdal:1971xh} 
  A.~A.~Migdal,
  ``On hadronic interactions at small distances,''
  Phys.\ Lett.\  {\bf 37B}, 98 (1971).
  doi:10.1016/0370-2693(71)90583-1
  %%CITATION = doi:10.1016/0370-2693(71)90583-1;%%
  %86 citations counted in INSPIRE as of 21 Jan 2020

%\cite{Polyakov:1970xd}
\bibitem{Polyakov:1970xd} 
  A.~M.~Polyakov,
  ``Conformal symmetry of critical fluctuations,''
  JETP Lett.\  {\bf 12}, 381 (1970)
  [Pisma Zh.\ Eksp.\ Teor.\ Fiz.\  {\bf 12}, 538 (1970)].
  %%CITATION = JTPLA,12,381;%%
  %355 citations counted in INSPIRE as of 21 Jan 2020


 %\cite{Polyakov:1974gs}
\bibitem{Polyakov:1974gs} 
  A.~M.~Polyakov,
  ``Nonhamiltonian approach to conformal quantum field theory,''
  Zh.\ Eksp.\ Teor.\ Fiz.\  {\bf 66}, 23 (1974)
  [Sov.\ Phys.\ JETP {\bf 39}, 9 (1974)].
  %%CITATION = ZETFA,66,23;%%
  %195 citations counted in INSPIRE as of 19 Sep 2017
       
%\cite{Ferrara:1973yt}
\bibitem{Ferrara:1973yt} 
  S.~Ferrara, A.~F.~Grillo and R.~Gatto,
  ``Tensor representations of conformal algebra and conformally covariant operator product expansion,''
  Annals Phys.\  {\bf 76}, 161 (1973).
  doi:10.1016/0003-4916(73)90446-6
  %%CITATION = doi:10.1016/0003-4916(73)90446-6;%%
  %219 citations counted in INSPIRE as of 19 Sep 2017
  
  

%\cite{Belavin:1984vu}
\bibitem{Belavin:1984vu} 
  A.~A.~Belavin, A.~M.~Polyakov and A.~B.~Zamolodchikov,
  ``Infinite Conformal Symmetry in Two-Dimensional Quantum Field Theory,''
  Nucl.\ Phys.\ B {\bf 241}, 333 (1984).
  doi:10.1016/0550-3213(84)90052-X
  %%CITATION = doi:10.1016/0550-3213(84)90052-X;%%
  %3276 citations counted in INSPIRE as of 19 Sep 2017
  
%\cite{Rattazzi:2008pe}
\bibitem{Rattazzi:2008pe} 
  R.~Rattazzi, V.~S.~Rychkov, E.~Tonni and A.~Vichi,
  ``Bounding scalar operator dimensions in 4D CFT,''
  JHEP {\bf 0812}, 031 (2008)
  doi:10.1088/1126-6708/2008/12/031
  [arXiv:0807.0004 [hep-th]].
  %%CITATION = doi:10.1088/1126-6708/2008/12/031;%%
  %359 citations counted in INSPIRE as of 19 Sep 2017

%\cite{Poland:2018epd}
\bibitem{Poland:2018epd} 
  D.~Poland, S.~Rychkov and A.~Vichi,
  ``The Conformal Bootstrap: Theory, Numerical Techniques, and Applications,''
  Rev.\ Mod.\ Phys.\  {\bf 91}, 015002 (2019)
  doi:10.1103/RevModPhys.91.015002
  [arXiv:1805.04405 [hep-th]].
  %%CITATION = doi:10.1103/RevModPhys.91.015002;%%
  %147 citations counted in INSPIRE as of 21 Jan 2020
  %\cite{El-Showk:2014dwa}
  
%\cite{ElShowk:2012ht}
\bibitem{ElShowk:2012ht}
  S.~El-Showk, M.~F.~Paulos, D.~Poland, S.~Rychkov, D.~Simmons-Duffin and A.~Vichi,
  ``Solving the 3D Ising Model with the Conformal Bootstrap,''
  Phys.\ Rev.\ D {\bf 86} (2012) 025022
  doi:10.1103/PhysRevD.86.025022
  [arXiv:1203.6064 [hep-th]].
  %%CITATION = doi:10.1103/PhysRevD.86.025022;%%
  %414 citations counted in INSPIRE as of 21 Jan 2020
  
\bibitem{El-Showk:2014dwa} 
  S.~El-Showk, M.~F.~Paulos, D.~Poland, S.~Rychkov, D.~Simmons-Duffin and A.~Vichi,
  ``Solving the 3d Ising Model with the Conformal Bootstrap II. c-Minimization and Precise Critical Exponents,''
  J.\ Stat.\ Phys.\  {\bf 157}, 869 (2014)
  doi:10.1007/s10955-014-1042-7
  [arXiv:1403.4545 [hep-th]].
  %%CITATION = doi:10.1007/s10955-014-1042-7;%%
  %272 citations counted in INSPIRE as of 21 Jan 2020

%\cite{Kos:2015mba}
\bibitem{Kos:2015mba} 
  F.~Kos, D.~Poland, D.~Simmons-Duffin and A.~Vichi,
  ``Bootstrapping the O(N) Archipelago,''
  JHEP {\bf 1511}, 106 (2015)
  doi:10.1007/JHEP11(2015)106
  [arXiv:1504.07997 [hep-th]].
  %%CITATION = doi:10.1007/JHEP11(2015)106;%%
  %138 citations counted in INSPIRE as of 21 Jan 2020
  
%\cite{Kos:2014bka}
\bibitem{Kos:2014bka} 
  F.~Kos, D.~Poland and D.~Simmons-Duffin,
  ``Bootstrapping Mixed Correlators in the 3D Ising Model,''
  JHEP {\bf 1411}, 109 (2014)
  doi:10.1007/JHEP11(2014)109
  [arXiv:1406.4858 [hep-th]].
  %%CITATION = doi:10.1007/JHEP11(2014)109;%%
  %211 citations counted in INSPIRE as of 21 Jan 2020
  
%\cite{Kos:2016ysd}
\bibitem{Kos:2016ysd} 
  F.~Kos, D.~Poland, D.~Simmons-Duffin and A.~Vichi,
  ``Precision Islands in the Ising and $O(N)$ Models,''
  JHEP {\bf 1608}, 036 (2016)
  doi:10.1007/JHEP08(2016)036
  [arXiv:1603.04436 [hep-th]].
  %%CITATION = doi:10.1007/JHEP08(2016)036;%%
  %63 citations counted in INSPIRE as of 19 Sep 2017

%\cite{Chester:2019ifh}
\bibitem{Chester:2019ifh} 
  S.~M.~Chester, W.~Landry, J.~Liu, D.~Poland, D.~Simmons-Duffin, N.~Su and A.~Vichi,
  ``Carving out OPE space and precise $O(2)$ model critical exponents,''
  arXiv:1912.03324 [hep-th].
  %%CITATION = ARXIV:1912.03324;%%
  %2 citations counted in INSPIRE as of 21 Jan 2020



%\cite{McKane:1984eq}
\bibitem{McKane:1984eq} 
  A.~J.~McKane and D.~J.~Wallace,
  ``Instanton Calculations Using Dimensional Regularization,''
  J.\ Phys.\ A {\bf 11}, 2285 (1978).
  doi:10.1088/0305-4470/11/11/013
  %%CITATION = doi:10.1088/0305-4470/11/11/013;%%
  %18 citations counted in INSPIRE as of 19 Feb 2020

  A.~J.~McKane, D.~J.~Wallace and O.~F.~de Alcantara Bonfim,
  ``Nonperturbative Renormalization Using Dimensional Regularization: Applications To The Epsilon Expansion,''
  J.\ Phys.\ A {\bf 17}, 1861 (1984).
  doi:10.1088/0305-4470/17/9/021
  %%CITATION = doi:10.1088/0305-4470/17/9/021;%%
  %7 citations counted in INSPIRE as of 23 Jan 2020


%\cite{Fei:2014yja}
\bibitem{Fei:2014yja} 
  L.~Fei, S.~Giombi and I.~R.~Klebanov,
  ``Critical $O(N)$ models in $6-\epsilon$ dimensions,''
  Phys.\ Rev.\ D {\bf 90}, no. 2, 025018 (2014)
  doi:10.1103/PhysRevD.90.025018
  [arXiv:1404.1094 [hep-th]].
  %%CITATION = doi:10.1103/PhysRevD.90.025018;%%
  %62 citations counted in INSPIRE as of 18 Sep 2017

%\cite{Fei:2014xta}
\bibitem{Fei:2014xta} 
  L.~Fei, S.~Giombi, I.~R.~Klebanov and G.~Tarnopolsky,
  ``Three loop analysis of the critical O(N) models in 6-$\epsilon$ dimensions,''
  Phys.\ Rev.\ D {\bf 91}, no. 4, 045011 (2015)
  doi:10.1103/PhysRevD.91.045011
  [arXiv:1411.1099 [hep-th]].
  %%CITATION = doi:10.1103/PhysRevD.91.045011;%%
  %42 citations counted in INSPIRE as of 18 Sep 2017


%\cite{Gracey:2015tta}
\bibitem{Gracey:2015tta} 
  J.~A.~Gracey,
  ``Four loop renormalization of $\phi^3$ theory in six dimensions,''
  Phys.\ Rev.\ D {\bf 92}, no. 2, 025012 (2015)
  doi:10.1103/PhysRevD.92.025012
  [arXiv:1506.03357 [hep-th]].
  %%CITATION = doi:10.1103/PhysRevD.92.025012;%%
  %35 citations counted in INSPIRE as of 18 Sep 2017
%\cite{Percacci:2014tfa}

%\cite{Giombi:2019upv}
\bibitem{Giombi:2019upv} 
  S.~Giombi, R.~Huang, I.~R.~Klebanov, S.~S.~Pufu and G.~Tarnopolsky,
  ``The $O(N)$ Model in $ {4 < d < 6} $ : Instantons and Complex CFTs,''
  arXiv:1910.02462 [hep-th].
  %%CITATION = ARXIV:1910.02462;%%
  %2 citations counted in INSPIRE as of 21 Jan 2020 
  


%\cite{Bae:2014hia}
\bibitem{Bae:2014hia} 
  J.~B.~Bae and S.~J.~Rey,
  ``Conformal Bootstrap Approach to O(N) Fixed Points in Five Dimensions,''
  arXiv:1412.6549 [hep-th].
  %%CITATION = ARXIV:1412.6549;%%
  %32 citations counted in INSPIRE as of 18 Sep 2017
 
%\cite{Chester:2014gqa}
\bibitem{Chester:2014gqa} 
  S.~M.~Chester, S.~S.~Pufu and R.~Yacoby,
  ``Bootstrapping $O(N)$ vector models in 4 $< d <$ 6,''
  Phys.\ Rev.\ D {\bf 91}, no. 8, 086014 (2015)
  doi:10.1103/PhysRevD.91.086014
  [arXiv:1412.7746 [hep-th]].
  %%CITATION = doi:10.1103/PhysRevD.91.086014;%%
  %49 citations counted in INSPIRE as of 18 Sep 2017

%\cite{Li:2016wdp}
\bibitem{Li:2016wdp} 
  Z.~Li and N.~Su,
  ``Bootstrapping Mixed Correlators in the Five Dimensional Critical O(N) Models,''
  JHEP {\bf 1704}, 098 (2017)
  doi:10.1007/JHEP04(2017)098
  [arXiv:1607.07077 [hep-th]].
  %%CITATION = doi:10.1007/JHEP04(2017)098;%%
  %12 citations counted in INSPIRE as of 18 Sep 2017
  



    
%\cite{El-Showk:2013nia}
\bibitem{El-Showk:2013nia} 
  S.~El-Showk, M.~Paulos, D.~Poland, S.~Rychkov, D.~Simmons-Duffin and A.~Vichi,
  ``Conformal Field Theories in Fractional Dimensions,''
  Phys.\ Rev.\ Lett.\  {\bf 112}, 141601 (2014)
  doi:10.1103/PhysRevLett.112.141601
  [arXiv:1309.5089 [hep-th]].
  %%CITATION = doi:10.1103/PhysRevLett.112.141601;%%
  %97 citations counted in INSPIRE as of 21 Jan 2020
 
%\cite{Hogervorst:2014rta,Hogervorst:2015akt}
\bibitem{Hogervorst:2014rta} 
  M.~Hogervorst, S.~Rychkov and B.~C.~van Rees,
  ``Truncated conformal space approach in d dimensions: A cheap alternative to lattice field theory?,''
  Phys.\ Rev.\ D {\bf 91}, 025005 (2015)
  doi:10.1103/PhysRevD.91.025005
  [arXiv:1409.1581 [hep-th]].
  %%CITATION = doi:10.1103/PhysRevD.91.025005;%%
  %74 citations counted in INSPIRE as of 30 Mar 2020
 
%\cite{Hogervorst:2015akt}
\bibitem{Hogervorst:2015akt} 
  M.~Hogervorst, S.~Rychkov and B.~C.~van Rees,
  ``Unitarity violation at the Wilson-Fisher fixed point in 4-$\epsilon$ dimensions,''
  Phys.\ Rev.\ D {\bf 93}, no. 12, 125025 (2016)
  doi:10.1103/PhysRevD.93.125025
  [arXiv:1512.00013 [hep-th]].
  %%CITATION = doi:10.1103/PhysRevD.93.125025;%%
  %51 citations counted in INSPIRE as of 30 Mar 2020   
 
\bibitem{Gliozzi:2013ysa} 
  F.~Gliozzi,
  ``More constraining conformal bootstrap,''
  Phys.\ Rev.\ Lett.\  {\bf 111}, 161602 (2013)
  doi:10.1103/PhysRevLett.111.161602
  [arXiv:1307.3111 [hep-th]].
  %%CITATION = doi:10.1103/PhysRevLett.111.161602;%%
  %74 citations counted in INSPIRE as of 19 Sep 2017
  
 \bibitem{Simmons-Duffin:2015qma} 
  D.~Simmons-Duffin,
  ``A Semidefinite Program Solver for the Conformal Bootstrap,''
  JHEP {\bf 1506}, 174 (2015)
  doi:10.1007/JHEP06(2015)174
  [arXiv:1502.02033 [hep-th]].
  %%CITATION = doi:10.1007/JHEP06(2015)174;%%
  %91 citations counted in INSPIRE as of 04 Sep 2017
%\cite{Heemskerk:2009pn}

 

%\cite{Gliozzi:2014jsa}
\bibitem{Gliozzi:2014jsa} 
  F.~Gliozzi and A.~Rago,
  ``Critical exponents of the 3d Ising and related models from Conformal Bootstrap,''
  JHEP {\bf 1410}, 042 (2014)
  doi:10.1007/JHEP10(2014)042
  [arXiv:1403.6003 [hep-th]].
  %%CITATION = doi:10.1007/JHEP10(2014)042;%%
  %64 citations counted in INSPIRE as of 19 Sep 2017
 

 
 
 %\cite{Heemskerk:2009pn}
\bibitem{Heemskerk:2009pn} 
  I.~Heemskerk, J.~Penedones, J.~Polchinski and J.~Sully,
  ``Holography from Conformal Field Theory,''
  JHEP {\bf 0910}, 079 (2009)
  doi:10.1088/1126-6708/2009/10/079
  [arXiv:0907.0151 [hep-th]].
  %%CITATION = doi:10.1088/1126-6708/2009/10/079;%%
  %264 citations counted in INSPIRE as of 09 Sep 2017


   
%\cite{Gorbenko:2018ncu}
\bibitem{Gorbenko:2018ncu} 
  V.~Gorbenko, S.~Rychkov and B.~Zan,
  ``Walking, Weak first-order transitions, and Complex CFTs,''
  JHEP {\bf 1810}, 108 (2018)
  doi:10.1007/JHEP10(2018)108
  [arXiv:1807.11512 [hep-th]].
  %%CITATION = doi:10.1007/JHEP10(2018)108;%%
  %41 citations counted in INSPIRE as of 23 Jan 2020  
  
  V.~Gorbenko, S.~Rychkov and B.~Zan,
  ``Walking, Weak first-order transitions, and Complex CFTs II. Two-dimensional Potts model at $Q>4$,''
  SciPost Phys.\  {\bf 5}, no. 5, 050 (2018)
  doi:10.21468/SciPostPhys.5.5.050
  [arXiv:1808.04380 [hep-th]].
  %%CITATION = doi:10.21468/SciPostPhys.5.5.050;%%
  %22 citations counted in INSPIRE as of 26 Jan 2020
  
%\cite{Dyson:1952tj}
\bibitem{Dyson:1952tj} 
  F.~J.~Dyson,
  ``Divergence of perturbation theory in quantum electrodynamics,''
  Phys.\ Rev.\  {\bf 85}, 631 (1952).
  doi:10.1103/PhysRev.85.631
  %%CITATION = doi:10.1103/PhysRev.85.631;%%
  %522 citations counted in INSPIRE as of 22 Jan 2020
  
 %\cite{Simon:1970mc}
\bibitem{Simon:1970mc} 
  B.~Simon and A.~Dicke,
  ``Coupling constant analyticity for the anharmonic oscillator,''
  Annals Phys.\  {\bf 58}, 76 (1970).
  doi:10.1016/0003-4916(70)90240-X
  %%CITATION = doi:10.1016/0003-4916(70)90240-X;%%
  %145 citations counted in INSPIRE as of 22 Jan 2020 
  
  J.~J.~Loeffel, A.~Martin, B.~Simon and A.~S.~Wightman,
  ``Pade approximants and the anharmonic oscillator,''
  Phys.\ Lett.\  {\bf 30B}, 656 (1969).
  doi:10.1016/0370-2693(69)90087-2
  %%CITATION = doi:10.1016/0370-2693(69)90087-2;%%
  %104 citations counted in INSPIRE as of 22 Jan 2020
  
%\cite{Vainshtein:1964dw}
\bibitem{Vainshtein:1964dw} 
  A.~I.~Vainshtein,
  ``Decaying systems and divergence of the series of perturbation theory,'' Novosibirsk Report, December 1964, reprinted in Russian, with an English translation by M. Shifman, in these Proceedings of QCD2002/ArkadyFest.
  
%\cite{Dymarsky:2005uh}
\bibitem{Dymarsky:2005uh} 
  A.~Dymarsky, I.~R.~Klebanov and R.~Roiban,
  ``Perturbative search for fixed lines in large N gauge theories,''
  JHEP {\bf 0508}, 011 (2005)
  doi:10.1088/1126-6708/2005/08/011
  [hep-th/0505099].
  %%CITATION = doi:10.1088/1126-6708/2005/08/011;%%
  %80 citations counted in INSPIRE as of 23 Jan 2020
  %\cite{Pomoni:2008de}
  
\bibitem{Pomoni:2008de} 
  E.~Pomoni and L.~Rastelli,
  ``Large N Field Theory and AdS Tachyons,''
  JHEP {\bf 0904}, 020 (2009)
  doi:10.1088/1126-6708/2009/04/020
  [arXiv:0805.2261 [hep-th]].
  %%CITATION = doi:10.1088/1126-6708/2009/04/020;%%
  %56 citations counted in INSPIRE as of 23 Jan 2020  
  
%\cite{Kaplan:2009kr}
\bibitem{Kaplan:2009kr} 
  D.~B.~Kaplan, J.~W.~Lee, D.~T.~Son and M.~A.~Stephanov,
  ``Conformality Lost,''
  Phys.\ Rev.\ D {\bf 80}, 125005 (2009)
  doi:10.1103/PhysRevD.80.125005
  [arXiv:0905.4752 [hep-th]].
  %%CITATION = doi:10.1103/PhysRevD.80.125005;%%
 
  %257 citations counted in INSPIRE as of 23 Jan 2020

%\cite{Giombi:2015haa}
\bibitem{Giombi:2015haa} 
  S.~Giombi, I.~R.~Klebanov and G.~Tarnopolsky,
  ``Conformal QED$_d$, $F$-Theorem and the $\epsilon$ Expansion,''
  J.\ Phys.\ A {\bf 49}, no. 13, 135403 (2016)
  doi:10.1088/1751-8113/49/13/135403
  [arXiv:1508.06354 [hep-th]].
  %%CITATION = doi:10.1088/1751-8113/49/13/135403;%%
  %60 citations counted in INSPIRE as of 23 Jan 2020

%\cite{Gracey:2015xmw}
\bibitem{Gracey:2015xmw}
J.~Gracey,
%``Six dimensional QCD at two loops,''
Phys. Rev. D \textbf{93}, no.2, 025025 (2016)
doi:10.1103/PhysRevD.93.025025
[arXiv:1512.04443 [hep-th]].
%25 citations counted in INSPIRE as of 30 May 2020
%\cite{Gliozzi:2017hni}
 %\cite{Gliozzi:2016ysv}
\bibitem{Gliozzi:2016ysv}
F.~Gliozzi, A.~Guerrieri, A.~C.~Petkou and C.~Wen,
%``Generalized Wilson-Fisher Critical Points from the Conformal Operator Product Expansion,''
Phys. Rev. Lett. \textbf{118}, no.6, 061601 (2017)
doi:10.1103/PhysRevLett.118.061601
[arXiv:1611.10344 [hep-th]].
%34 citations counted in INSPIRE as of 30 May 2020 
  \bibitem{Gliozzi:2017hni}
F.~Gliozzi, A.~L.~Guerrieri, A.~C.~Petkou and C.~Wen,
%``The analytic structure of conformal blocks and the generalized Wilson-Fisher fixed points,''
JHEP \textbf{04}, 056 (2017)
doi:10.1007/JHEP04(2017)056
[arXiv:1702.03938 [hep-th]].
%40 citations counted in INSPIRE as of 30 May 2020


\end{thebibliography}
\end{document}